\documentclass[5p]{elsarticle}
\pdfoutput=1 

\usepackage{dcolumn}
\usepackage{graphicx}
\usepackage{color}
\usepackage{amsmath}
\usepackage{amssymb}
\usepackage{amsthm}
\usepackage{latexsym}
\usepackage{bm}
\usepackage{multirow}
\usepackage{cancel}
\usepackage{hyperref}

\usepackage{tabularx}
\newcolumntype{Z}{>{\centering\arraybackslash}X}

\newcommand*{\gev}{{\rm GeV}}
\newcommand*{\tev}{{\rm TeV}}

\newcommand*{\beq}{\begin{equation}}
\newcommand*{\eeq}{\end{equation}}
\newcommand*{\bea}{\begin{eqnarray}}
\newcommand*{\eea}{\end{eqnarray}}

\def\h{\ensuremath{h}}
\def\A{\ensuremath{A}}

\newcommand*{\mh}{\ensuremath{m_h}}
\newcommand*{\mA}{\ensuremath{m_A}}

\def\Hboson{\ensuremath{H}}
\def\antibar#1{\ensuremath{#1\bar{#1}}}
\def\ttbar{\antibar{t}}

\def\ttH{\ensuremath{\ttbar\Hboson}}
\def\ggH{\ensuremath{gg\mathrm{F}}}
\def\VBFH{\ensuremath{\mathrm{VBF}}}
\def\VH{\ensuremath{V\Hboson}}

\newcommand{\rf}{\ensuremath{\mu_{\ggH+\ttH}^{\gamma\gamma}}}
\newcommand{\rv}{\ensuremath{\mu_{\VBFH+\VH}^{\gamma\gamma}}}
\newcommand{\muzz}{\ensuremath{\mu^{4\ell}}}

\newcommand*{\hpm}{\ensuremath{H^\pm}}
\newcommand*{\hob}{\ensuremath{H_{\rm obs}}}

\newcommand*{\mhpm}{\ensuremath{m_{H^\pm}}}

\newcommand*{\half}{\ensuremath{\frac{1}{2}}}
\newcommand*{\tanb}{\ensuremath{\tan\beta}}

\journal{Physics Letters B}

\begin{document}

\begin{frontmatter}
\title{Electroweak production of light scalar-pseudoscalar pairs\\ from extended Higgs sectors}

\author[upp]{Rikard Enberg}
\author[upp,man]{William Klemm}
\author[soton]{Stefano Moretti}
\author[kias]{Shoaib Munir}

\address[upp]{Department of Physics and Astronomy, Uppsala University, 
Box 516, SE-751 20 Uppsala, Sweden}
\address[man]{School of Physics \& Astronomy, University of Manchester, 
Manchester M13 9PL, UK}
\address[soton]{School of Physics \& Astronomy,
University of Southampton, Southampton SO17 1BJ, UK}
\address[kias]{School of Physics, Korea Institute for Advanced Study,
Seoul 130-722, Republic of Korea}

\begin{abstract}
In models with extended Higgs sectors, it is possible that the Higgs boson discovered at the LHC is not the lightest one.
We show that in a realistic model (the Type I 2-Higgs Doublet Model), when the sum of the masses of a light scalar and a pseudoscalar ($h$ and $A$) is smaller than the $Z$ boson mass, the Electroweak (EW) production of an $hA$ pair
can dominate over QCD production by orders of magnitude, a fact not previously highlighted. 
This is because in the $gg$-initiated process, $hA$ production via a resonant $Z$ in the $s$-channel is prohibited according to the Landau-Yang theorem, which is not the case for the $q\bar{q}$-initiated process. We explore the parameter space of the model to highlight regions giving such $hA$ solutions while being consistent with all constraints from collider searches, $b$-physics and EW precision data. We also single out a few benchmark points to discuss their salient features, including the $hA$ search channels that can be exploited at Run II of the LHC.
\end{abstract}

\end{frontmatter}

\section{Introduction}
Most models for physics beyond the Standard Model (SM)
predict extended Higgs sectors, with additional Higgs (pseudo)scalars.
Two-Higgs Doublet Models (2HDMs), which contain two Higgs 
doublets $\phi_1$ and $\phi_2$ (see\ \cite{Branco:2011iw}
for a review), are among the simplest non-trivial
extensions of the SM. 
The Higgs sector of a CP-conserving 2HDM contains 
three neutral Higgs bosons, two scalars and a pseudoscalar 
($h$, $H$, with $m_h<m_H$, and $A$, respectively), and a charged 
pair \hpm. One of the two CP-even Higgs bosons must 
have properties consistent with the observed 
$125$\ \gev\ state\
\cite{Aad:2012tfa,Chatrchyan:2012ufa,Chatrchyan:2012jja}, \hob.
At the Large Hadron Collider (LHC), the neutral Higgs bosons 
of a 2HDM can be 
produced both singly, dominantly via gluon fusion, and 
in identical or mixed pairs. 
We discuss here a scenario in which the $h$ and $A$ states of 
the Type-I 2HDM (2HDM-I),\footnote{In the Type I model, 
all fermions get mass from Yukawa
couplings to only one of the doublets, see below.} with 
masses satisfying $m_h+m_A< M_Z$, can pass the present experimental 
constraints from the Large Electron Positron (LEP) collider, the Tevatron 
and the LHC, with the heavier $H$ state being identified with \hob.

\begin{figure}[t!]
\begin{center}
    \includegraphics[scale=0.65]{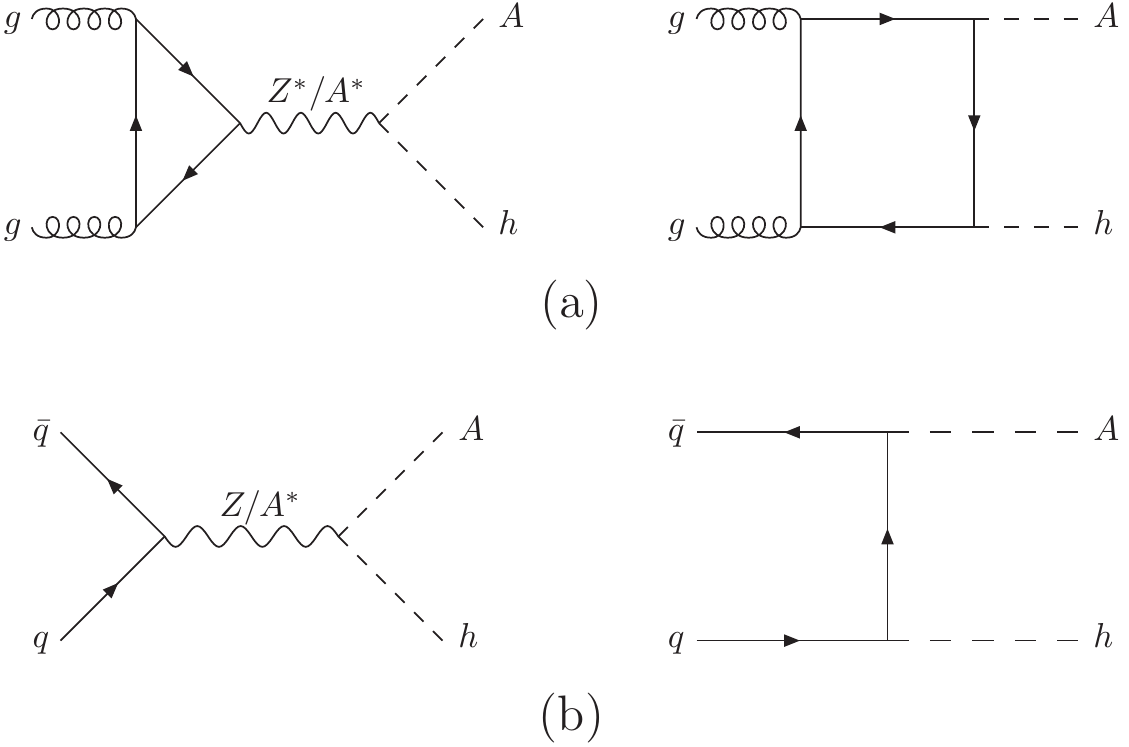}
    \caption{Diagrams contributing to (a) QCD production and (b) EW production of the $hA$ pair.}
    \label{fig:graphs}
\end{center}
\end{figure}

The LHC is a hadron collider that can yield collisions with
very small momentum fraction $x$ of the scattered partons and very large 
squared momentum transfer $Q^2$. Because the proton
has a large gluon density at small $x$,  
one would hope to initiate $Z$ production from gluon-gluon ($gg$) scattering 
(see the left diagram of Fig.~\ref{fig:graphs}a), with the 
$hA$ final state produced from $Z$ decay.
However, owing to the Landau-Yang theorem~\cite{Landau:1948kw,Yang:1950rg}, $gg$ can only scatter 
via a $Z$ if it is non-resonant
(i.e., off-shell, denoted by $Z^*$)\ \cite{Moretti:2014rka}.
This leads to a much depleted cross section 
for the $hA$ signal and, additionally, to the inability of using 
$Z$ mass reconstruction from the invariant mass of the $hA$ (visible) 
decay products for suppressing backgrounds. In the case of the 
tree-level quark-antiquark ($q\bar q$)-initiated process, however, 
the $Z$ boson can be produced on-shell (left diagram of 
Fig.~\ref{fig:graphs}b). 
The $hA$ final state can also be produced from double Higgs-strahlung 
off heavy quarks (i.e., $b$- and $t$-quarks), at the one-loop level
(right diagram of Fig.~\ref{fig:graphs}a) and at the tree level 
(right diagram of Fig.~\ref{fig:graphs}b), in the case of $gg$ 
and $q\bar q$ collisions, respectively.

It is the purpose of this Letter to highlight the hitherto 
neglected predominance of the $q\bar q$-initiated tree-level production 
of a light $hA$ pair at the LHC with respect to the $gg$-initiated
one-loop production in a Type-I 2HDM. (See Ref. \cite{Dawson:1998py} for higher order QCD corrections to the corresponding diagrams.) We additionally outline the region of the 2HDM-I 
parameter space where the former can be accessed above and beyond the yield 
of the latter and present benchmark points to serve as a guideline for 
probing this production process at the current LHC run.

\section{Model, parameter scan and constraints}
In general, in a 2HDM, 
depending on how the 
two doublets couple to fermions, Flavor Changing Neutral Currents
(FCNCs) can be mediated by (pseudo)scalars at
the tree level. The requirement of vanishing FCNCs thus puts very strong
restrictions on the coupling matrices. The simplest way 
to avoid large FCNCs is to impose a $Z_2$ symmetry so that each 
type of fermion only couples to one of the
doublets (``natural flavor 
conservation'')\ \cite{Glashow:1976nt,Paschos:1976ay}. 
There are four basic ways of assigning the $Z_2$ charges, and here we consider the case where only the doublet $\phi_2$ couples to all fermions, known as the Type I model.
The Higgs potential for the CP-conserving 2HDM-I is written as
\bea
V  &=& m_{11}^2\phi_1^\dagger\phi_1+ m_{22}^2\phi_2^\dagger\phi_2
-[m_{12}^2\phi_1^\dagger\phi_2+ \, \text{h.c.} ] \nonumber \\
&+&\half\lambda_1(\phi_1^\dagger\phi_1)^2
+\half\lambda_2(\phi_2^\dagger\phi_2)^2
+\lambda_3(\phi_1^\dagger\phi_1)(\phi_2^\dagger\phi_2) \nonumber\\ 
&+&\lambda_4(\phi_1^\dagger\phi_2)(\phi_2^\dagger\phi_1)
+ [\half\lambda_5(\phi_1^\dagger\phi_2)^2 + 
\text{h.c.}], 
\eea
which is invariant under the symmetry $\phi_1 \to -\phi_1$ up to the 
soft breaking term proportional to $m_{12}^ 2$. 
Through the minimization conditions of the potential, $m_{11}^2$ and $m_{22}^2$ 
can be traded for the vacuum expectation values,  
 $v_1$ and $v_2$, of the two Higgs fields and the tree-level mass relations 
allow the quartic couplings $\lambda_{1-5}$ to be
substituted by the four physical Higgs boson masses and the neutral
sector term {$s_{\beta-\alpha}$ (short for $\sin(\beta-\alpha)$, with the angle $\beta$ defined through $\tanb=v_2/v_1$)}, where $\alpha$ mixes the CP-even Higgs states.

In order to test the consistency of solutions with 
$m_h+m_A < M_Z$ in the 2HDM-I with the most crucial and 
relevant theoretical and experimental constraints
 (listed further below), we performed a scan of its 
parameter space\footnote{Note that a similar region of parameter space was captured by Ref.~\cite{Bernon:2015wef}} using 
2HDMC-v1.7.0\ \cite{Eriksson:2009ws}. The (randomly) scanned ranges 
of the free parameters (with $m_H=$ 125\ \gev) are given 
in the second column of Tab.~\ref{tab:params}.  
Because only a select region of the parameter space is allowed by current constraints, we used the distributions resulting from this initial scan to determine the most relevant parameter ranges, which we focused on in a second scan, shown in the rightmost column of Tab.~\ref{tab:params}.

\begin{table}[tb]
\begin{center}
	\begin{tabularx}{\columnwidth}{l X l}
\hline 
Parameter & Initial range & Refined range \\
\hline
		\mh\ (\gev) & (10, 80) & (10, $2M_Z/3$) \\
		\mA\ (\gev) & (10, $M_Z-m_h$) & ($m_h/2$, $M_Z-m_h$) \\
		\mhpm (\gev) & (90, 500) & (90, 150) \\
		$s_{\beta-\alpha}$ & ($-1$, 1) & ($-0.25$, 0) \\
		$m_{12}^2$ (\gev$^2$) & (0, $m_{\A}^2\sin\beta\cos\beta$) & (0, $m_{\A}^2\sin\beta\cos\beta$) \\
		\tanb & (2, 25) & $(-0.95,-1.1)/s_{\beta-\alpha}$ \\
\hline
	\end{tabularx}
	\caption{2HDM-I parameters and their scanned ranges.}
	\label{tab:params}
\end{center}
\end{table}

During the scan, each sampled model point was subjected 
to the following conditions: \\
\noindent -- Unitarity, perturbativity, and vacuum 
stability enforced through the default 2HDMC method. \\
\noindent -- Consistency at 95\% Confidence Level (CL) with 
the experimental measurements of 
the oblique parameters $S$, $T$ and $U$, again, 
calculated  by 2HDMC. 
We compare these to the fit values\ \cite{Agashe:2014kda}, $S=0.00 \pm 0.08$ and $T=0.05 \pm 0.07$,  
in an ellipse with a correlation of 90\%. 
All points further satisfy $U=0.05\pm 0.10$.\\
\noindent -- Satisfaction of the 95\% CL limits on  $b$-physics 
observables calculated with the public code SuperIso-v3.4\ \cite{Mahmoudi:2008tp}.\\
\noindent -- Consistency with the $Z$ width measurement from LEP, $\Gamma_Z = 2.4952\pm 0.0023$\
\gev\ \cite{Agashe:2014kda}. The partial width $\Gamma(Z\to h A)$ was
 required to fall within the $2\sigma$ experimental
uncertainty of the measurement. \\
\noindent -- Consistency of the mass and signal rates of $H$ with the 
LHC data on  \hob. The combined 68\% CL results from ATLAS and CMS 
for the most sensitive channels are\ \cite{Aad:2015zhl}: 
$\rf=1.15^{+0.28}_{-0.25}$, $\rv=1.17^{+0.58}_{-0.53}$, $\muzz=1.40^{+0.30}_{-0.25}$.
We required that the equivalent quantities, calculated with
HiggsSignals-v1.3.2\ \cite{Bechtle:2013xfa}, satisfy these
measurements at 95\% CL, assuming Gaussian uncertainties. \\
\noindent -- Consistency of all Higgs states with the direct search constraints 
from LEP, Tevatron, and LHC at the $95\%$ CL tested using the public tool HiggsBounds-v4.3.1\ 
\cite{Bechtle:2008jh,Bechtle:2011sb,Bechtle:2013gu,Bechtle:2013wla}.

The points were also required
to satisfy some additional constraints from LEP and LHC
that have not (yet) been implemented in HiggsBounds.
Consistency with the combined LEP
$\hpm$ searches in the 2HDM-I\ \cite{Abbiendi:2013hk} was ensured
by requiring that $m_{H^\pm} > 90~\gev$. The LEP-II constraints on
$e^+e^-\to\gamma\gamma b\bar{b}$\ \cite{Abdallah:2003xf} were also
taken into account. While these constraints are mass dependent, we
conservatively required $\cos^2(\beta-\alpha){\rm
  BR}(h\to\gamma\gamma){\rm BR}(A\to b\bar{b}) < 0.02$. Moreover, 
the results of the $\mu\mu\tau\tau$ final state studies performed by ATLAS\
\cite{Aad:2015oqa} as well as of the $\tau \tau \tau \tau$\ \cite{CMS:2015iga},
$\mu\mu\tau\tau$\ \cite{CMS:2016cel} and $\mu\mu b\bar{b}$\ \cite{CMS:2016cqw}
analyses from CMS were tested against.


\section{Scan results}
From the output of our initial scan, we noticed that the LHC
observation of a very SM-like $\hob$ pushes the model towards the
alignment limit, $s_{\beta-\alpha} \to 0$. Additionally, strong
constraints from LEP searches lead to suppressed $h$/$A$ couplings to
fermions,\footnote{In the 2HDM-I, the couplings of $h$ and $A$ to fermions go as $g_{hf\bar{f}}\sim \cos\alpha/sin\beta$ and $g_{Af\bar{f}}\sim\pm\cot\beta$.} producing a strong correlation $s_{\beta-\alpha}\approx
-1/\tan\beta$.
We also find that a relatively light charged Higgs ($m_{H^\pm}\lesssim 120~\gev$) is necessary, as a charged Higgs mass too far separated from $m_h$ or $m_A$ results in large contributions to the $T$-parameter.\footnote{This requirement of a light charged Higgs prevents us from finding similar points in Type-II models, where a higher $m_{H^\pm}$ is required by $B$-physics constraints.}
Existing searches for charged Higgs bosons in this mass range typically focus on production from top decays followed by charged Higgs boson decays to either $\tau\nu$ or $cs$. For the points selected by the scan, these branching ratios typically fall below the percent level, in many cases by several orders of magnitude, with maximal values of BR$(t\to H^+ b) \lesssim 0.04$, BR$(H^+ \to \tau^+ \nu_\tau)\lesssim 0.01$, and BR$(H^+ \to c\bar{s}) \lesssim 6 \times 10^{-3}$. This places them well below existing constraints, including recent LHC results\ \cite{Aad:2014kga,Khachatryan:2015qxa,Khachatryan:2015uua} not yet included in HiggsBounds. Instead of the standard decays, the low masses of $h$ and $A$ in the scenario considered here allow the $H^\pm$ to decay dominantly in the $W^* h$ or $W^* A$ channels (with the respective branching ratios alternatively near unity), which have not yet been examined at the LHC.\footnote{These decay modes of the $H^\pm$ will be discussed further in\ \cite{smallmass_charged}.}

\begin{figure}[tbp]
\begin{center}
    \includegraphics[scale=0.5]{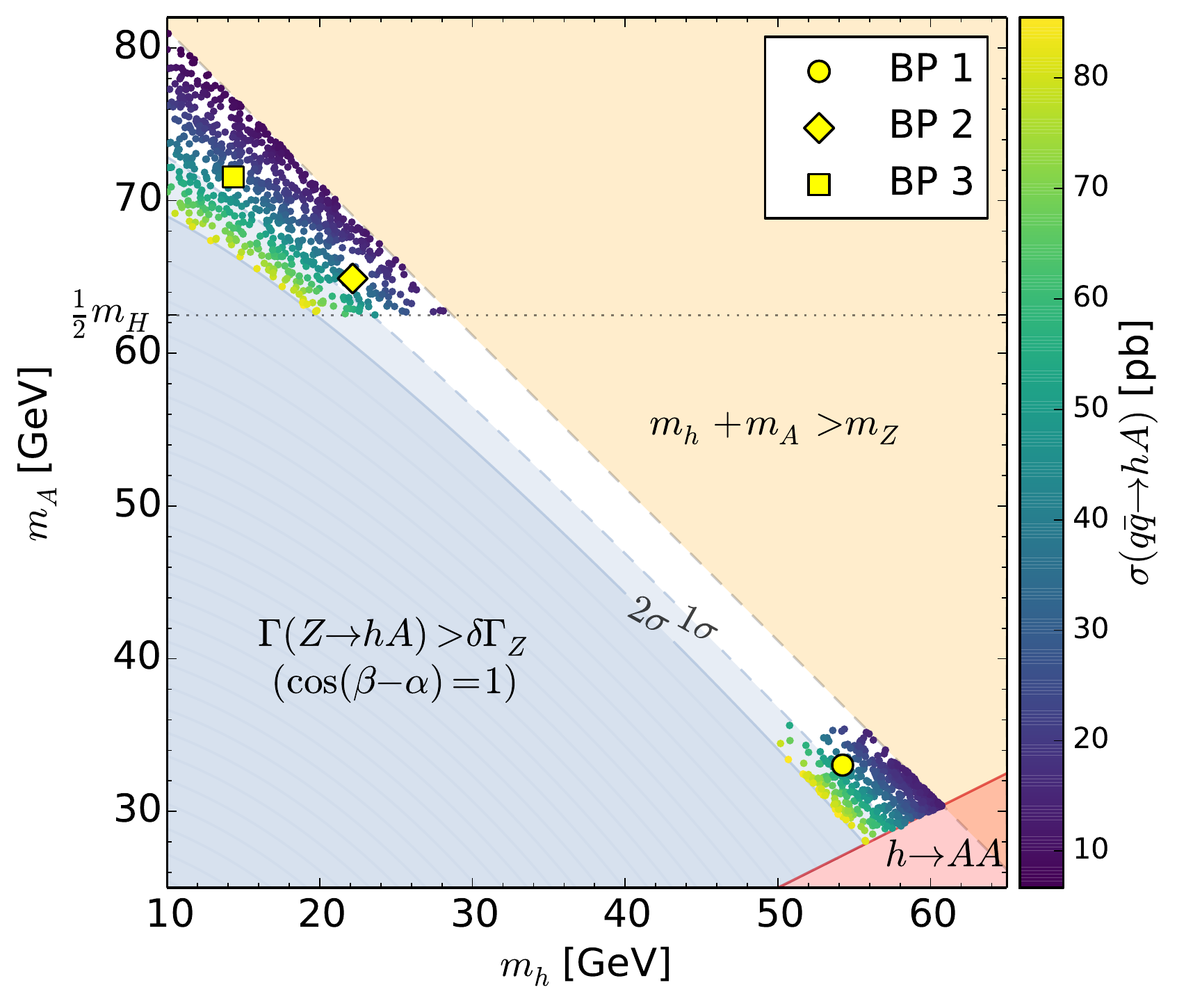}
    \caption{Constraints and accepted points in the $(m_h,m_A)$ plane.  
Shaded areas: Red -- $m_\h > 2m_\A$, allowing $\h\to\A\A$ decays; Blue
-- theoretical prediction of the $Z\to\h\A$ partial width exceeds
experimental uncertainty at the $1\sigma$ (lighter) and $2\sigma$
(darker) levels, in the limit $\cos(\beta-\alpha)=1$; Orange -- $m_h +
m_A$ above the $m_Z$ threshold, not considered in this study. The
color map corresponds to the total cross section for the $q\bar{q}\to
hA$ process at $\sqrt{s}=13~\tev$, and the three benchmark points 
have been highlighted in yellow. }
    \label{fig:ZWidth}
\end{center}
\end{figure}

Numerous constraints restrict the possible masses of $h$ and $A$.
In Fig.~\ref{fig:ZWidth} we show the
points passing all the constraints mentioned above in the $(m_h,m_A)$
plane. Because the $hAZ$ coupling is maximized in the favored
$s_{\beta-\alpha}\to 0$ limit, the constraint from $\Gamma_Z$, the
$1\sigma$ and $2\sigma$ contours for which are also shown, is
particularly severe. We note two distinct regions with 
a large density of points in the figure. 
The region near the top left corner corresponds to the $m_A > m_h$ (heavier
$A$) scenario. This region cuts off sharply at $m_A=m_H/2$ due to 
the possibility of the $H\to A A$ decay arising, which potentially
leads to a suppression of the signal strengths for the SM-like $H$ (for
the 2HDM-I scenarios we consider, these signal strengths are always
below 1 to begin with).  This possibility can be avoided with a
sufficiently suppressed $HAA$ coupling, as a result of which 
additional points satisfying all constraints appear in the 
region corresponding to the $m_h > m_A$ (heavier $h$) scenario near
the lower right corner of the figure.
When $m_h > 2m_A$, the $h \to A A$ decay channel 
opens up, and the model is severely constrained by LEP searches for processes such as
$e^+ e^- \to h A \to (AA)A \to (b\bar{b}b\bar{b})b\bar{b}$\
\cite{Schael:2006cr}. Consequently, we did not find acceptable points with $m_h > 2m_A$.

The color map in Fig.~\ref{fig:ZWidth} depicts the total cross section for the
$q\bar{q}'\to hA$ process, which evidently grows larger as one moves
away from the diagonal and $m_h + m_A$ gets smaller. For calculating
this cross section, we used the 2HDMC model\ \cite{Eriksson:2009ws} 
with MadGraph5\_aMC@NLO\ \cite{Alwall:2014hca}, considering both 
4- ($q=u,d,c,s$) and 5- ($q=u,d,c,s,b$) flavor schemes. The 5-flavor
scheme predictions differ by less than 3\% from those of the 4-flavor
one due to the small $b$-quark couplings. Also highlighted
in the figure are the three Benchmark Points (BPs) selected
to demonstrate the typical characteristics of the interesting parameter
space regions. These BPs will be discussed in detail later.

\section{EW vs.\ QCD production}
\begin{figure}[tbp]
\begin{center}
    \includegraphics[scale=0.45]{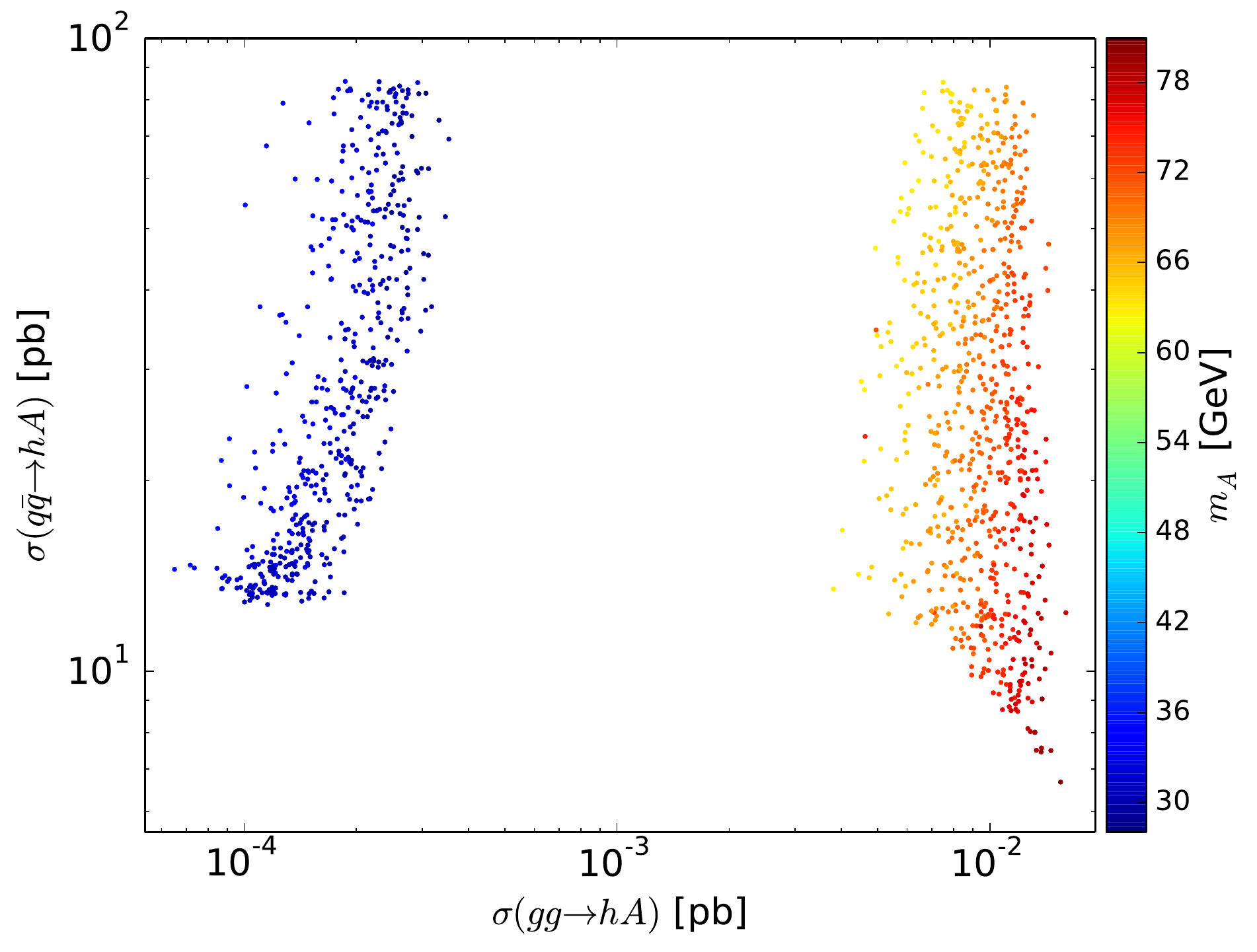}
    \caption{Cross sections for $q\bar{q}$- vs. $gg$-initiated $hA$
      production at the LHC with $\sqrt{s}=13~\tev$, for points
      satisfying all the constraints described in the text. The color map indicates $m_A$.}
    \label{fig:qqvsgg}
\end{center}
\end{figure}

In order to be able to compare the
relative strengths of the $q\bar{q}'\to hA$ production mode and the
$gg\to hA$ mode, we also calculated the cross section for the latter
for each point using codes developed with MadGraph5\_aMC@NLO\
\cite{Alwall:2014hca} for Higgs pair production\
\cite{Hespel:2014sla}. The comparison is shown in
Fig.~\ref{fig:qqvsgg}, where one notices
that the maximal cross section achievable for QCD production is about
three orders of magnitude smaller than that for EW production, which
can reach as high as $\sim 90$\,pb. 
Also, for the points shown, while the maximal cross section for EW
production is consistent across
the two $(m_h,m_A)$ regions, which can be distinguished through the
color map in $m_A$, QCD production clearly prefers the heavier $A$ scenario. 

\section{Benchmarks}
The input parameters for the three BPs shown in
Fig.~\ref{fig:ZWidth} are given in Tab.~\ref{tab:BP} along with the
corresponding cross sections in the two $hA$ production channels analyzed. 
BP1 corresponds to the heavier $h$ scenario while BP2 and BP3 
correspond to the heavier $A$ scenario.
\begin{table*}[t]
\begin{center}
	\begin{tabularx}{2\columnwidth}{Z Z Z Z Z Z Z Z Z}
        \hline
		BP & \mh  & \mA & \mhpm & $s_{\beta-\alpha}$ & $m_{12}^2$ & \tanb & $\sigma (q\bar{q})$ & $\sigma (gg)$ \\\hline
1 & 54.2 & 33.0 & 95.9 & $-0.12$ & 118.3 & 9.1 & 41.2 & $1.5 \times 10^{-4}$\\ 
2 & 22.2 & 64.9 & 101.5 & $-0.05$ & 10.6 & 22.1 & 34.4 & $7.2 \times 10^{-3}$\\ 
3 & 14.3 & 71.6 & 107.2 & $-0.06$ & 2.9 & 16.3 & 31.6 & $1.1 \times 10^{-2}$\\
        \hline
	\end{tabularx}
    \caption{Input parameters and parton-level cross sections (in pb) corresponding to the selected benchmark points. All masses are in GeV and for all points $m_H=125~\gev$.}
	\label{tab:BP}
\end{center}
\end{table*}

In Tab. \ref{tab:BR} we list the BRs of $h$ and $A$ in the most
important decay channels for each BP.  The allowed points in the
heavier $h$ scenario all have characteristics similar to BP1 -- a
highly fermiophobic $h$ which consequently decays dominantly to $Z^*
A$ and a light $A$ which decays primarily into pairs of third
generation fermions.\footnote{The $h\to Z^* A$ and $A\to Z^* h$ decays
  were previously discussed in a fermiophobic model in\
  \cite{Akeroyd:1998dt}.} The main signatures of interest would then
include $Z^*b\bar{b}b\bar{b}$, $Z^*b\bar{b}\tau\tau$, and
$Z^*\tau\tau\tau\tau$.  A similar situation is also possible in the
heavier $A$ scenario, as seen for BP2, where the roles of $A$ and $h$
are now reversed, but the most common final states remain the
same. Unlike $A$, however, the light $h$ can also decay dominantly to two
photons (due to contribution from $W^\pm$ loops, which is missing in the
$A\to\gamma\gamma$ decay), thus opening up the possibility of 
$Ah \to (Z^*h) h \to Z^*\gamma\gamma\gamma\gamma$ or 
$Z^*\gamma\gamma b \bar{b}$ decay chain for points like BP3.

\begin{table}[h!]
\begin{center}
	\begin{tabularx}{\columnwidth}{Z  Z Z Z Z  Z  Z Z Z}
    \hline
{} & \multicolumn{4}{c}{BR$(h\to ...)$ [\%]} & {} &
\multicolumn{3}{c}{BR$(A\to ...)$ [\%]} \\
BP & $Z^* A$ & $b\bar{b}$ & $\gamma\gamma$ & $\tau\tau$ & {} & $Z^* h$ & $b\bar{b}$ & $\tau\tau$ \\ \hline
1 & \textbf{94} & 5 & $<1$ & $<1$ & {} & 0 & \textbf{86} & 7\\ 
2 & 0 & \textbf{83} & 3 & 7 & {} & \textbf{86} & 12 & 1\\ 
3 & 0 & \textbf{60} & \textbf{24} & 7 & {} & \textbf{90} & 8 & 1\\
    \hline
	\end{tabularx}
    \caption{Dominant BRs [\%] of $h$ and $A$ for the BPs. BRs greater than 20\% are highlighted in bold.}
	\label{tab:BR}
\end{center}
\end{table}


\section{Concluding remarks}
In summary, we have shown that, even
when the most up-to-date theoretical and experimental constraints are
imposed, the 2HDM-I offers an intriguing phenomenological situation
wherein $m_h+m_A< m_Z$. This possibility is precluded in other 2HDM
Types. Such $hA$ pairs can be produced in $q\bar{q}$-annihilation 
via resonant $Z$ in the $s$-channel, unlike the case
of $gg$ fusion, where their production can only proceed via non-resonant 
$Z^*$, owing to the Landau-Yang theorem. As a consequence, at the
LHC Run II, the former would yield event rates up to four orders of
magnitude larger than the latter. Taking into account also the
double Higgs-strahlung production, the inclusive rates for the
$q\bar{q} \to hA$ process can be as large as tens of pb, and hence
amenable to experimental investigation and potential discovery 
by the LHC {\sl already at present}. 

Finally, to demonstrate their feasibility, we have
provided a few 2HDM-I parameter configurations producing distinctive 
$hA$ decay patterns. We look forward to the ATLAS and CMS experiments 
testing this hitherto neglected scenario against their data, as
establishing one or more of the potential $hA$ signatures discussed
here will provide not only a direct proof of a non-minimal Higgs
sector but also circumstantial evidence of a specific 2HDM structure.


\section*{Acknowledgments}
The work of RE and SMo is funded through the grant 
H2020-MSCA-RISE-2014 No. 645722 (NonMinimalHiggs). SMo is supported in part through the NExT Institute  and STFC Consolidated Grant ST/J000396/1. This work is also supported by the Swedish Research Council under contract 621-2011-5107.

\section*{References}
\bibliography{hA_electroweak}

\begin{thebibliography}{10}
\expandafter\ifx\csname url\endcsname\relax
  \def\url#1{\texttt{#1}}\fi
\expandafter\ifx\csname urlprefix\endcsname\relax\def\urlprefix{URL }\fi
\expandafter\ifx\csname href\endcsname\relax
  \def\href#1#2{#2} \def\path#1{#1}\fi

\bibitem{Branco:2011iw}
G.~Branco, P.~Ferreira, L.~Lavoura, M.~Rebelo, M.~Sher, et~al., {Theory and
  phenomenology of two-Higgs-doublet models}, Phys. Rept. 516 (2012) 1--102.
\newblock \href {http://arxiv.org/abs/1106.0034} {\path{arXiv:1106.0034}},
  \href {http://dx.doi.org/10.1016/j.physrep.2012.02.002}
  {\path{doi:10.1016/j.physrep.2012.02.002}}.

\bibitem{Aad:2012tfa}
G.~Aad, et~al., {Observation of a new particle in the search for the Standard
  Model Higgs boson with the ATLAS detector at the LHC}, Phys. Lett. B716
  (2012) 1--29.
\newblock \href {http://arxiv.org/abs/1207.7214} {\path{arXiv:1207.7214}},
  \href {http://dx.doi.org/10.1016/j.physletb.2012.08.020}
  {\path{doi:10.1016/j.physletb.2012.08.020}}.

\bibitem{Chatrchyan:2012ufa}
S.~Chatrchyan, et~al., {Observation of a new boson at a mass of 125 GeV with
  the CMS experiment at the LHC}, Phys. Lett. B716 (2012) 30--61.
\newblock \href {http://arxiv.org/abs/1207.7235} {\path{arXiv:1207.7235}},
  \href {http://dx.doi.org/10.1016/j.physletb.2012.08.021}
  {\path{doi:10.1016/j.physletb.2012.08.021}}.

\bibitem{Chatrchyan:2012jja}
S.~Chatrchyan, et~al., {Study of the Mass and Spin-Parity of the Higgs Boson
  Candidate Via Its Decays to Z Boson Pairs}, Phys. Rev. Lett. 110 (2013)
  081803.
\newblock \href {http://arxiv.org/abs/1212.6639} {\path{arXiv:1212.6639}},
  \href {http://dx.doi.org/10.1103/PhysRevLett.110.081803}
  {\path{doi:10.1103/PhysRevLett.110.081803}}.

\bibitem{Landau:1948kw}
L.~D. Landau, {On the angular momentum of a system of two photons}, Dokl. Akad.
  Nauk Ser. Fiz. 60~(2) (1948) 207--209.
\newblock \href {http://dx.doi.org/10.1016/B978-0-08-010586-4.50070-5}
  {\path{doi:10.1016/B978-0-08-010586-4.50070-5}}.

\bibitem{Yang:1950rg}
C.-N. Yang, {Selection Rules for the Dematerialization of a Particle Into Two
  Photons}, Phys. Rev. 77 (1950) 242--245.
\newblock \href {http://dx.doi.org/10.1103/PhysRev.77.242}
  {\path{doi:10.1103/PhysRev.77.242}}.

\bibitem{Moretti:2014rka}
S.~Moretti, {Variations on a Higgs theme}, Phys. Rev. D91~(1) (2015) 014012.
\newblock \href {http://arxiv.org/abs/1407.3511} {\path{arXiv:1407.3511}},
  \href {http://dx.doi.org/10.1103/PhysRevD.91.014012}
  {\path{doi:10.1103/PhysRevD.91.014012}}.

\bibitem{Dawson:1998py}
S.~Dawson, S.~Dittmaier, M.~Spira, {Neutral Higgs boson pair production at
  hadron colliders: QCD corrections}, Phys. Rev. D58 (1998) 115012.
\newblock \href {http://arxiv.org/abs/hep-ph/9805244}
  {\path{arXiv:hep-ph/9805244}}, \href
  {http://dx.doi.org/10.1103/PhysRevD.58.115012}
  {\path{doi:10.1103/PhysRevD.58.115012}}.

\bibitem{Glashow:1976nt}
S.~L. Glashow, S.~Weinberg, {Natural Conservation Laws for Neutral Currents},
  Phys. Rev. D15 (1977) 1958.
\newblock \href {http://dx.doi.org/10.1103/PhysRevD.15.1958}
  {\path{doi:10.1103/PhysRevD.15.1958}}.

\bibitem{Paschos:1976ay}
E.~Paschos, {Diagonal Neutral Currents}, Phys. Rev. D15 (1977) 1966.
\newblock \href {http://dx.doi.org/10.1103/PhysRevD.15.1966}
  {\path{doi:10.1103/PhysRevD.15.1966}}.

\bibitem{Bernon:2015wef}
J.~Bernon, J.~F. Gunion, H.~E. Haber, Y.~Jiang, S.~Kraml, {Scrutinizing the
  alignment limit in two-Higgs-doublet models. II. m$_H$=125  GeV}, Phys.
  Rev. D93~(3) (2016) 035027.
\newblock \href {http://arxiv.org/abs/1511.03682} {\path{arXiv:1511.03682}},
  \href {http://dx.doi.org/10.1103/PhysRevD.93.035027}
  {\path{doi:10.1103/PhysRevD.93.035027}}.

\bibitem{Eriksson:2009ws}
D.~Eriksson, J.~Rathsman, O.~St\aa{}l, {2HDMC: Two-Higgs-Doublet Model
  Calculator Physics and Manual}, Comput. Phys. Commun. 181 (2010) 189--205.
\newblock \href {http://arxiv.org/abs/0902.0851} {\path{arXiv:0902.0851}},
  \href {http://dx.doi.org/10.1016/j.cpc.2009.09.011}
  {\path{doi:10.1016/j.cpc.2009.09.011}}.

\bibitem{Agashe:2014kda}
K.~Olive, et~al., {Review of Particle Physics}, Chin. Phys. C38 (2014) 090001.
\newblock \href {http://dx.doi.org/10.1088/1674-1137/38/9/090001}
  {\path{doi:10.1088/1674-1137/38/9/090001}}.

\bibitem{Mahmoudi:2008tp}
F.~Mahmoudi, {SuperIso v2.3: A Program for calculating flavor physics
  observables in Supersymmetry}, Comput. Phys. Commun. 180 (2009) 1579--1613.
\newblock \href {http://arxiv.org/abs/0808.3144} {\path{arXiv:0808.3144}},
  \href {http://dx.doi.org/10.1016/j.cpc.2009.02.017}
  {\path{doi:10.1016/j.cpc.2009.02.017}}.

\bibitem{Aad:2015zhl}
G.~Aad, et~al., {Combined Measurement of the Higgs Boson Mass in $pp$
  Collisions at $\sqrt{s}=7$ and 8 TeV with the ATLAS and CMS Experiments},
  Phys. Rev. Lett. 114 (2015) 191803.
\newblock \href {http://arxiv.org/abs/1503.07589} {\path{arXiv:1503.07589}},
  \href {http://dx.doi.org/10.1103/PhysRevLett.114.191803}
  {\path{doi:10.1103/PhysRevLett.114.191803}}.

\bibitem{Bechtle:2013xfa}
P.~Bechtle, S.~Heinemeyer, O.~St\aa{}l, T.~Stefaniak, G.~Weiglein,
  {$HiggsSignals$: Confronting arbitrary Higgs sectors with measurements at the
  Tevatron and the LHC}, Eur. Phys. J. C74 (2014) 2711.
\newblock \href {http://arxiv.org/abs/1305.1933} {\path{arXiv:1305.1933}},
  \href {http://dx.doi.org/10.1140/epjc/s10052-013-2711-4}
  {\path{doi:10.1140/epjc/s10052-013-2711-4}}.

\bibitem{Bechtle:2008jh}
P.~Bechtle, O.~Brein, S.~Heinemeyer, G.~Weiglein, K.~E. Williams, {HiggsBounds:
  Confronting Arbitrary Higgs Sectors with Exclusion Bounds from LEP and the
  Tevatron}, Comput.Phys.Commun. 181 (2010) 138--167.
\newblock \href {http://arxiv.org/abs/0811.4169} {\path{arXiv:0811.4169}},
  \href {http://dx.doi.org/10.1016/j.cpc.2009.09.003}
  {\path{doi:10.1016/j.cpc.2009.09.003}}.

\bibitem{Bechtle:2011sb}
P.~Bechtle, O.~Brein, S.~Heinemeyer, G.~Weiglein, K.~E. Williams, {HiggsBounds
  2.0.0: Confronting Neutral and Charged Higgs Sector Predictions with
  Exclusion Bounds from LEP and the Tevatron}, Comput. Phys. Commun. 182 (2011)
  2605--2631.
\newblock \href {http://arxiv.org/abs/1102.1898} {\path{arXiv:1102.1898}},
  \href {http://dx.doi.org/10.1016/j.cpc.2011.07.015}
  {\path{doi:10.1016/j.cpc.2011.07.015}}.

\bibitem{Bechtle:2013gu}
P.~Bechtle, O.~Brein, S.~Heinemeyer, O.~St\aa{}l, T.~Stefaniak, et~al., {Recent
  Developments in HiggsBounds and a Preview of HiggsSignals}, PoS CHARGED2012
  (2012) 024.
\newblock \href {http://arxiv.org/abs/1301.2345} {\path{arXiv:1301.2345}}.

\bibitem{Bechtle:2013wla}
P.~Bechtle, O.~Brein, S.~Heinemeyer, O.~St\aa{}l, T.~Stefaniak, et~al.,
  {$\mathsf{HiggsBounds}-4$: Improved Tests of Extended Higgs Sectors against
  Exclusion Bounds from LEP, the Tevatron and the LHC}, Eur. Phys. J. C74
  (2014) 2693.
\newblock \href {http://arxiv.org/abs/1311.0055} {\path{arXiv:1311.0055}},
  \href {http://dx.doi.org/10.1140/epjc/s10052-013-2693-2}
  {\path{doi:10.1140/epjc/s10052-013-2693-2}}.

\bibitem{Abbiendi:2013hk}
G.~Abbiendi, et~al., {Search for Charged Higgs bosons: Combined Results Using
  LEP Data}, Eur. Phys. J. C73 (2013) 2463.
\newblock \href {http://arxiv.org/abs/1301.6065} {\path{arXiv:1301.6065}},
  \href {http://dx.doi.org/10.1140/epjc/s10052-013-2463-1}
  {\path{doi:10.1140/epjc/s10052-013-2463-1}}.

\bibitem{Abdallah:2003xf}
J.~Abdallah, et~al., {Search for fermiophobic Higgs bosons in final states with
  photons at LEP 2}, Eur. Phys. J. C35 (2004) 313--324.
\newblock \href {http://arxiv.org/abs/hep-ex/0406012}
  {\path{arXiv:hep-ex/0406012}}, \href
  {http://dx.doi.org/10.1140/epjc/s2004-01869-2}
  {\path{doi:10.1140/epjc/s2004-01869-2}}.

\bibitem{Aad:2015oqa}
G.~Aad, et~al., {Search for Higgs bosons decaying to $aa$ in the
  $\mu\mu\tau\tau$ final state in $pp$ collisions at $\sqrt{s} = $ 8 TeV with
  the ATLAS experiment}, Phys. Rev. D92~(5) (2015) 052002.
\newblock \href {http://arxiv.org/abs/1505.01609} {\path{arXiv:1505.01609}},
  \href {http://dx.doi.org/10.1103/PhysRevD.92.052002}
  {\path{doi:10.1103/PhysRevD.92.052002}}.

\bibitem{CMS:2015iga}
S.~Chatrchyan, et~al., {Search for Higgs Decays to New Light Bosons in Boosted
  Tau Final States}, Report No. CMS-PAS-HIG-14-022.

\bibitem{CMS:2016cel}
S.~Chatrchyan, et~al., {Search for exotic decays of the Higgs boson to a pair
  of new light bosons with two muon and two b jets in final states}, Report No.
  CMS-PAS-HIG-14-041.

\bibitem{CMS:2016cqw}
S.~Chatrchyan, et~al., {Search for the exotic decay of the Higgs boson to two
  light pseudoscalar bosons with two taus and two muons in the final state at
  $\sqrt{s} = 8$ TeV}, Report No. CMS-PAS-HIG-15-011.

\bibitem{Aad:2014kga}
G.~Aad, et~al., {Search for charged Higgs bosons decaying via $H^{\pm}
  \rightarrow \tau^{\pm}\nu$ in fully hadronic final states using $pp$
  collision data at $\sqrt{s} = 8$ TeV with the ATLAS detector}, JHEP 03 (2015)
  088.
\newblock \href {http://arxiv.org/abs/1412.6663} {\path{arXiv:1412.6663}},
  \href {http://dx.doi.org/10.1007/JHEP03(2015)088}
  {\path{doi:10.1007/JHEP03(2015)088}}.

\bibitem{Khachatryan:2015qxa}
V.~Khachatryan, et~al., {Search for a charged Higgs boson in pp collisions at $
  \sqrt{s}=8 $ TeV}, JHEP 11 (2015) 018.
\newblock \href {http://arxiv.org/abs/1508.07774} {\path{arXiv:1508.07774}},
  \href {http://dx.doi.org/10.1007/JHEP11(2015)018}
  {\path{doi:10.1007/JHEP11(2015)018}}.

\bibitem{Khachatryan:2015uua}
V.~Khachatryan, et~al., {Search for a light charged Higgs boson decaying to $
  \mathrm{c}\overline{\mathrm{s}} $ in pp collisions at $ \sqrt{s}=8 $ TeV},
  JHEP 12 (2015) 178.
\newblock \href {http://arxiv.org/abs/1510.04252} {\path{arXiv:1510.04252}},
  \href {http://dx.doi.org/10.1007/JHEP12(2015)178}
  {\path{doi:10.1007/JHEP12(2015)178}}.

\bibitem{smallmass_charged}
A.~Arhrib, R.~Benbrik, R.~Enberg, W.~Klemm, S.~Moretti, S.~Munir, {in
  progress}.

\bibitem{Schael:2006cr}
S.~Schael, et~al., {Search for neutral MSSM Higgs bosons at LEP}, Eur. Phys. J.
  C47 (2006) 547--587.
\newblock \href {http://arxiv.org/abs/hep-ex/0602042}
  {\path{arXiv:hep-ex/0602042}}, \href
  {http://dx.doi.org/10.1140/epjc/s2006-02569-7}
  {\path{doi:10.1140/epjc/s2006-02569-7}}.

\bibitem{Alwall:2014hca}
J.~Alwall, R.~Frederix, S.~Frixione, V.~Hirschi, F.~Maltoni, O.~Mattelaer,
  H.~S. Shao, T.~Stelzer, P.~Torrielli, M.~Zaro, {The automated computation of
  tree-level and next-to-leading order differential cross sections, and their
  matching to parton shower simulations}, JHEP 07 (2014) 079.
\newblock \href {http://arxiv.org/abs/1405.0301} {\path{arXiv:1405.0301}},
  \href {http://dx.doi.org/10.1007/JHEP07(2014)079}
  {\path{doi:10.1007/JHEP07(2014)079}}.

\bibitem{Hespel:2014sla}
B.~Hespel, D.~Lopez-Val, E.~Vryonidou, {Higgs pair production via gluon fusion
  in the Two-Higgs-Doublet Model}, JHEP 09 (2014) 124.
\newblock \href {http://arxiv.org/abs/1407.0281} {\path{arXiv:1407.0281}},
  \href {http://dx.doi.org/10.1007/JHEP09(2014)124}
  {\path{doi:10.1007/JHEP09(2014)124}}.

\bibitem{Akeroyd:1998dt}
A.~G. Akeroyd, {Three body decays of Higgs bosons at LEP-2 and application to a
  hidden fermiophobic Higgs}, Nucl. Phys. B544 (1999) 557--575.
\newblock \href {http://arxiv.org/abs/hep-ph/9806337}
  {\path{arXiv:hep-ph/9806337}}, \href
  {http://dx.doi.org/10.1016/S0550-3213(98)00845-1}
  {\path{doi:10.1016/S0550-3213(98)00845-1}}.

\end{thebibliography}
\bibliographystyle{elsarticle-num}

\end{document}